 \documentclass[final,3p,times,twocolumn]{elsarticle}

\usepackage{amssymb}
\usepackage{multirow}
\usepackage{amsmath}
\usepackage{url}
\usepackage{xcolor}
\usepackage[normalem]{ulem}

\journal{Astroparticle Physics}

\begin{document}

\begin{frontmatter}



\title{Dual MeV Gamma-Ray and Dark Matter Observatory - GRAMS Project}


\author[SLAC]{T. Aramaki} 
\ead{tsuguo@slac.stanford.edu}
\author[SLAC]{P.H. Adrian}

\author[COLUMBIA]{Georgia Karagiorgi}
\author[Tokyo]{Hirokazu Odaka}

\address[SLAC]{SLAC National Accelerator Laboratory/Kavli Institute for Particle Astrophysics and Cosmology, Menlo Park, CA 94025, USA}
\address[COLUMBIA]{Department of Physics, Columbia University, New York, NY 10027, USA}
\address[Tokyo]{Department of Physics, University of Tokyo, Tokyo 113-0033, Japan}


\begin{abstract}
GRAMS (Gamma-Ray and AntiMatter Survey) is a novel project that can simultaneously target both astrophysical observations with MeV gamma rays and an indirect dark matter search with antimatter. The GRAMS instrument is designed with a cost-effective, large-scale LArTPC (Liquid Argon Time Projection Chamber) detector surrounded by plastic scintillators. The astrophysical observations at MeV energies have not yet been well-explored (the so-called ``MeV-gap'') and GRAMS can improve the sensitivity by more than an order of magnitude compared to previous experiments. While primarily focusing on MeV gamma-ray observations, GRAMS is also optimized for cosmic ray antimatter surveys to indirectly search for dark matter. In particular, low-energy antideuterons will provide an essentially background-free dark matter signature. GRAMS will be a next generation experiment beyond the current GAPS (General AntiParticle Spectrometer) project for antimatter survey.

\end{abstract}

\begin{keyword}
MeV gamma rays, dark matter, antimatter, antiprotons, antideuterons, antiheliums, GRAMS, LArTPC, GAPS, AMS-02, Fermi GCE

\end{keyword}

\end{frontmatter}


\section{Introduction} \label{sec:intro}

In recent astrophysical observations, the Large Area Telescope on the Fermi Gamma-ray Space Telescope (Fermi-LAT) and the Nuclear Spectroscopic Telescope Array (NuSTAR) opened new windows to survey astrophysical phenomena in the energy domains for hard X-rays (up to 80 keV) and high-energy gamma rays (above 20 MeV), respectively \citep{Atwood2009,Harrison2013}. However, gamma rays in the MeV energy range have not yet been well-explored (the so-called ``MeV gap''). COMPTEL (The Imaging COMPton TELescope) produced the first catalogue of MeV sources, but only approximately 30 objects have been detected \citep{Schonfelder2000}. Gamma-ray line astronomy, particularly in the MeV energy region, is the key to understanding nucleosynthesis processes by direct observations of nuclear emission lines. The radioactive isotopes can serve as chronometers and tracers while providing information on the physical conditions during nucleosynthesis. Moreover, energetic particle acceleration can be studied through MeV gamma-ray observations since the transition from thermal to non-thermal physical processes can occur in the MeV energy region. These phenomena can be seen in relativistic flows generated in stellar mass black holes, supermassive black holes in active galactic nuclei and various types of neutron stars such as radio pulsars and magnetars \citep{Longair2011}. In addition, multi-messenger astronomy is an important means to understand astrophysical objects including transient phenomena. MeV gamma rays may be produced in association with gravitational waves from neutron star mergers \citep{Abbott2017}.

\begin{figure*}[t!]
\begin{center} 
\includegraphics*[width=12cm]{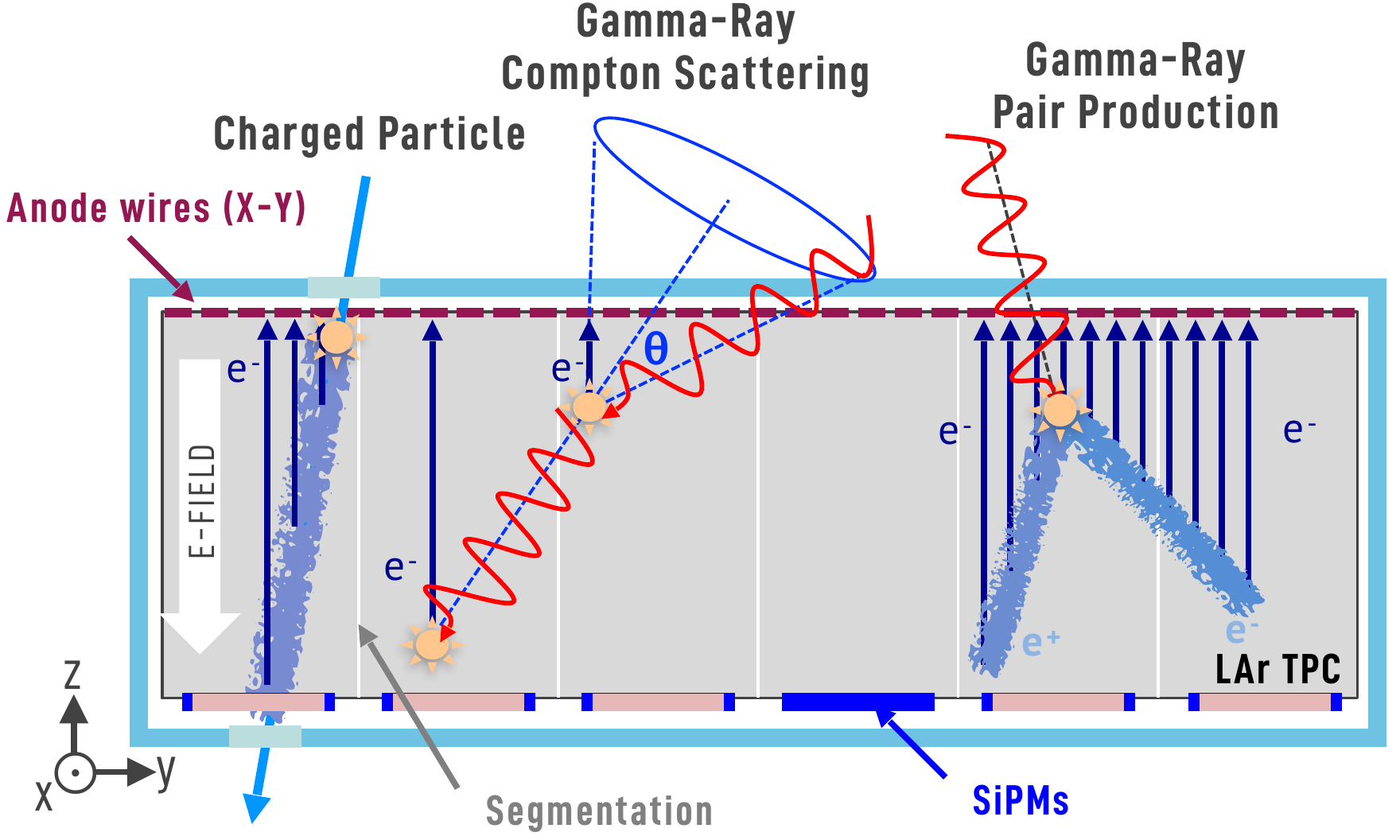}
\end{center}
\caption{Detection concept for charged particles and gamma rays (Compton scattering and pair-production).}
\label{fig:TPC}
\end{figure*}

Astrophysical observations of gravitational lensing, the Bullet Cluster, and galaxy rotation curves have indicated the existence of dark matter since the 1960s. The nature and origin of dark matter are still unknown, and many theories and experiments have been proposed to understand dark matter. The recent results of Fermi-LAT and Alpha Magnetic Spectrometer (AMS-02) suggested possible dark matter signatures in gamma-ray observations and antiproton measurements respectively \citep{Calore2015,Daylan2016,Abazajian2016,Cui2016,Korsmeier2018}. These results are, however, in strong tension with Fermi dwarf spheroidal galaxies observations \citep{Ackermann2015,Ackermann2017,Korsmeier2018}. The GAPS project, a current generation experiment, could elucidate the tension by measuring antiprotons and antideuterons produced by dark matter annihilation or decay. The GRAMS project, a next generation experiment, could have an extended sensitivity to antideuterons, which could potentially allow us to obtain crude spectrum information for antideuterons (see Section \ref{sec:AM_Sensitivity}).


GRAMS is the first project to simultaneously target both astrophysical observations with MeV gamma rays and an indirect dark matter search with antimatter using a Liquid Argon Time Projection Chamber (LArTPC) detector. Liquid noble gas detectors have been used for gamma-ray astronomy and underground dark matter search experiments. The LXeGRIT (The Liquid Xenon Gamma-Ray Imaging Telescope) experiment successfully operated a LXeTPC detector during a balloon flight \citep{Aprile1998,Aprile2000,Curioni2007}. The performance of the LArTPC detector, in particular, has been significantly improved over the past decade for neutrino and dark matter search experiments \citep{Chepel2013}. The large-scale LArTPC detector in GRAMS, unlike current and previous experiments with semiconductors or scintillation crystals, can offer enhanced sensitivities to gamma rays and antiparticles at very modest cost as described in the following sections.



\section{Instrument Design} \label{sec:instrument}

\begin{figure}[b!]
\begin{center} 
\includegraphics*[width=7.5cm]{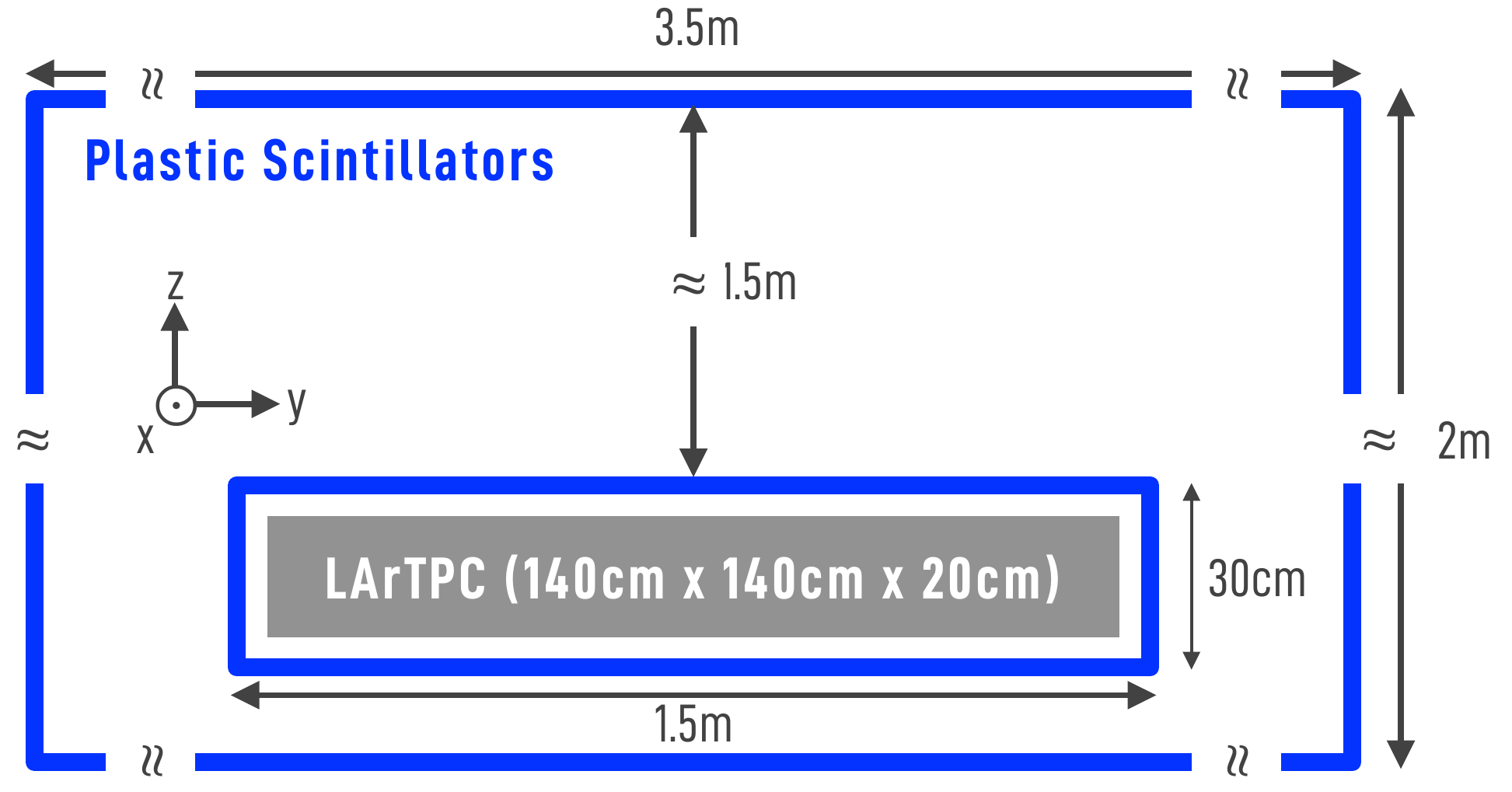}
\end{center}
\caption{GRAMS detector: LArTPC surrounded by plastic scintillators}
\label{fig:Det}
\end{figure}

The GRAMS instrumentation is composed of a LArTPC detector surrounded by two layers of plastic scintillators. The LArTPC detector in GRAMS will approximately be 140 $\times$ 140 $\times$ 20 cm while the overall instrument size will approximately be 3.5 $\times$ 3.5 $\times$ 2 m (see Figure \ref{fig:Det}). The preliminary mass estimate indicates that the payload size is compatible with the Long-Duration Balloon (LDB) flight. Unlike semiconductor or scintillation detectors, the GRAMS detector is cost-effective, which allows a large-scale detector considering that argon is both plentiful and low-cost. The LArTPC detector works as a Compton camera and a calorimeter for MeV gamma-ray observations as well as a particle tracker for antimatter measurements. The plastic scintillators will veto the incoming charged particles for MeV gamma-ray observations while triggering the incoming particles for antimatter detection by measuring the time-of-flight (TOF) between the outer and inner scintillator layers. The LArTPC detector and the readout electronics will be cooled down to $\sim$85 K while the plastic scintillators will be operated at ambient temperature.

\begin{figure*}[t!]
\begin{center} 
\includegraphics*[width=13.5cm]{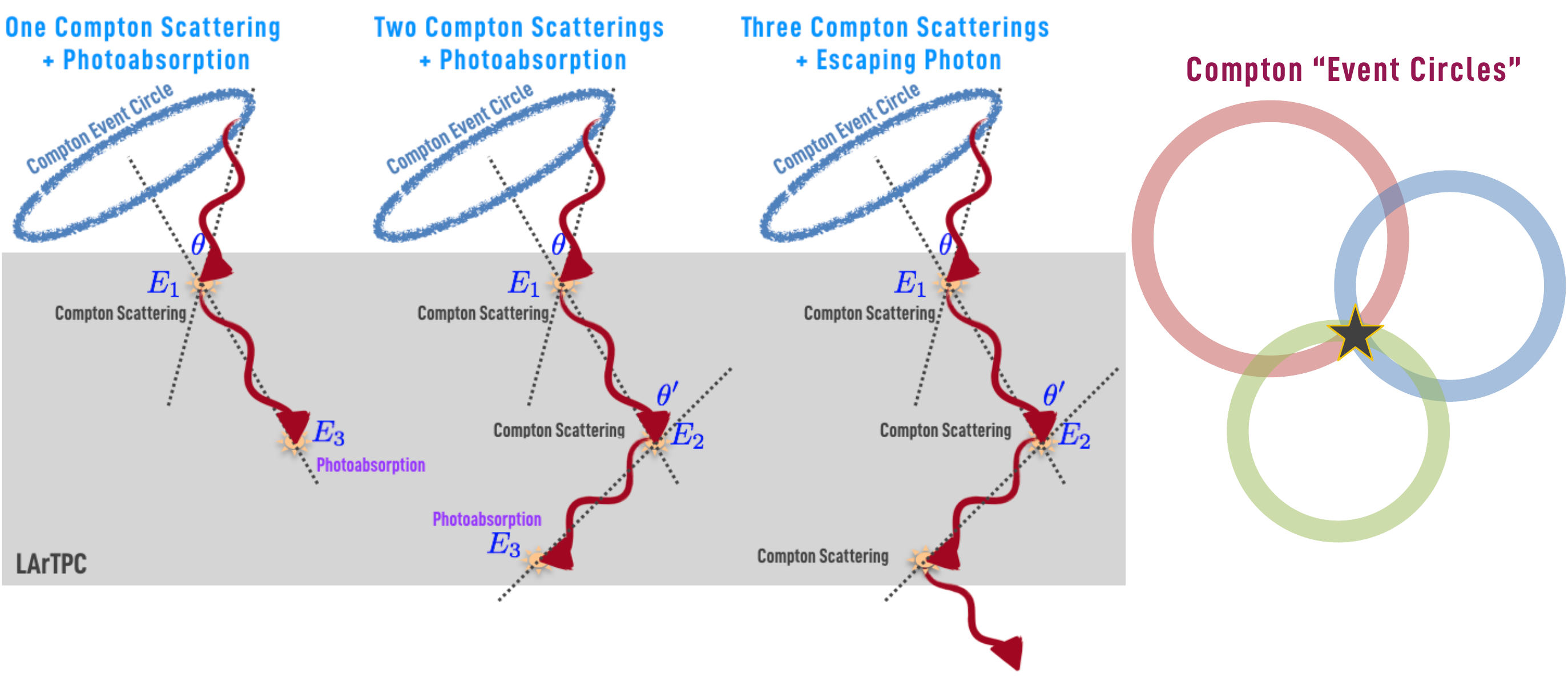}
\end{center}
\caption{Compton-scattering events in GRAMS. Three or more Compton ``Event Circles'' provide the direction of the gamma-ray source.}
\label{fig:Compton}
\end{figure*}

Particles entering the LArTPC detector excite and ionize argon atoms, producing scintillation light and ionization electrons. The scintillation light measured by Silicon Photo-Multipliers (SiPMs) is used for triggering and timing of the event. The ionization electrons drift in an applied electric field and are collected at anode planes with a $\sim$2 mm pitch of wires. The signals induced on the anode wires provide the x and y coordinates of the event, while the drift time of the ionization electrons gives the z position. The ability to reconstruct three-dimensional space points without a multi-layered design is one of the key advantages for GRAMS compared to other gamma-ray detectors with semiconductors or scintillation crystals. To reduce the coincident background during the finite collection time of the ionization signal, the LArTPC drift volume is segmented into ``cells'' with PTFE sheets (see Figure \ref{fig:TPC}). The detector can be further upgraded for a future satellite mission that focuses on the gamma-ray observations. The overall instrumental size can be $\sim$1.5m $\times$ 1.5m $\times$ 30cm without the outer scintillator layer while using a fine pitch ($\sim$0.2~mm) of anode wires to track Compton-scattered electrons. We can also add a calorimeter around the detector, such as a CZT (Cadmium Zinc Telluride) detector, similar to a future satellite mission e-ASTROGAM \citep{De2017}, in order to measure the escaping gamma rays and pair-produced electrons and positions. We are currently working towards the development of a prototype detector, and the details of the readout electronics and the telemetry system will be determined in the future.

\section{MeV Gamma-Ray Observation} \label{sec:GR}

\subsection{Detection Concept}

The MeV gamma-ray survey requires accurate reconstruction of the photon energy and direction. For energies above the electron-positron pair-production threshold ($E > 2m_eC^2$), GRAMS uses the precise tracking capability of the LArTPC to reconstruct the momentum of the electron-positron to determine the incident gamma ray. At lower energies, where gamma rays preferably undergo a Compton scattering and a photo-absorption, GRAMS relies on accurately determining the position and energy of the Compton electron and photo-absorption. Reconstruction of the incident gamma-ray energy $E$ and cone angle $\theta$ can be estimated by the Compton equation \citep{Kamae1987,Dogan1990}. A gamma ray may undergo a number of Compton scatterings before being photo-absorbed or even escape the sensitive volume of the detector (see Figure \ref{fig:Compton}). For one or two Compton scatterings followed by photo-absorption, the energy and cone angle can be calculated as below. 

\begin{align*}
E &= E_1 + E_2 +E_3\\
cos\theta &= 1-m_e c^2 \left( \frac{1}{E_2+E_3} - \frac{1}{E_1+E_2+E_3} \right)\\
cos\theta' &= 1-m_e c^2 \left( \frac{1}{E_3} - \frac{1}{E_2+E_3} \right)
\end{align*}
where $m_e$ is the electron mass, $E_1$, $E_2$, and $E_3$ are the deposited energies by the Compton scatterings and photo-absorption, and $\theta$ and $\theta'$ are the first and the second Compton scattering angles. For a single Compton scattering followed by a photo-absorption, the above simplifies since there is no $\theta'$ and $E_2 = 0$. For an event with three or more Compton scatterings, the angle and energy can be reconstructed as below. 

\begin{align*}
E &= E_1 + E_2 + E_3' \\
cos\theta &= 1-m_e c^2 \left( \frac{1}{E_2+E_3'} - \frac{1}{E_1+E_2+E_3'} \right)\\
E_3' &= -\frac{E_2}{2}+\sqrt{\frac{E_2^2}{4}+\frac{E_2m_e c^2}{1-cos\theta'}}
\end{align*}
where $E_3'$ is the energy of the gamma ray just after the second Compton scattering. The gamma ray after three or more Compton scatterings may escape from the detector. In each case, the cone angle $\theta$ gives a Compton ``Event Circle'' for each event, and the overlap of three or more ``Event Circles'' pinpoints the direction of the gamma-ray source (see Figure \ref{fig:Compton}).

\subsection{Effective Area}

\begin{figure}[h]
\begin{center} 
\includegraphics*[width=8.0cm]{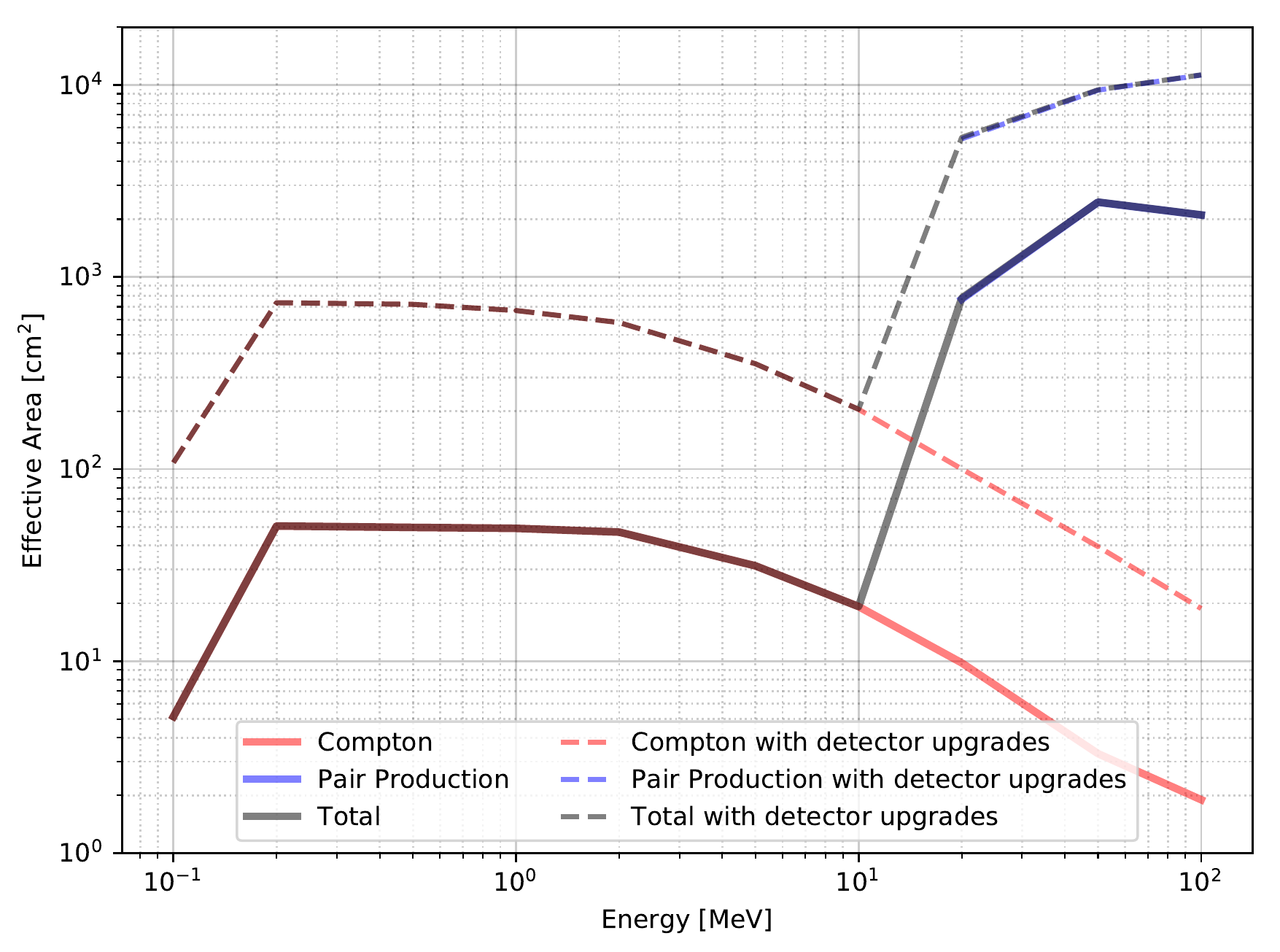}
\end{center}
\caption{Effective areas for Compton scattering and pair-production events in GRAMS (solid lines). The dashed lines represent the effective areas with detector upgrades.}
\label{fig:EA}
\end{figure}

The effective area for gamma rays with energy of $0.1$ MeV $< E < 100$ MeV was estimated using a GEANT4 simulation \citep{Agostinelli2003}. For reliable event reconstruction, events with one or two Compton scatterings followed by photo-absorption or three Compton scatterings inside the LArTPC detector were considered. In order to improve the angular resolution, a set of event selections was also applied; Compton scatterings must be spatially separated by $\geq 10$ (2) cm, and pair-produced electrons and positrons must stop inside the sensitive volume and leave tracks $\geq 2$ (0.4) cm long (with detector upgrades). Figure \ref{fig:EA} shows the effective areas for Compton scattering and pair-production events in GRAMS (solid lines). The dashed lines represent the effective areas with detector upgrades. 

\subsection{Angular and Energy Resolution}

\begin{figure}[b!]
\begin{center} 
\includegraphics*[width=8.0cm]{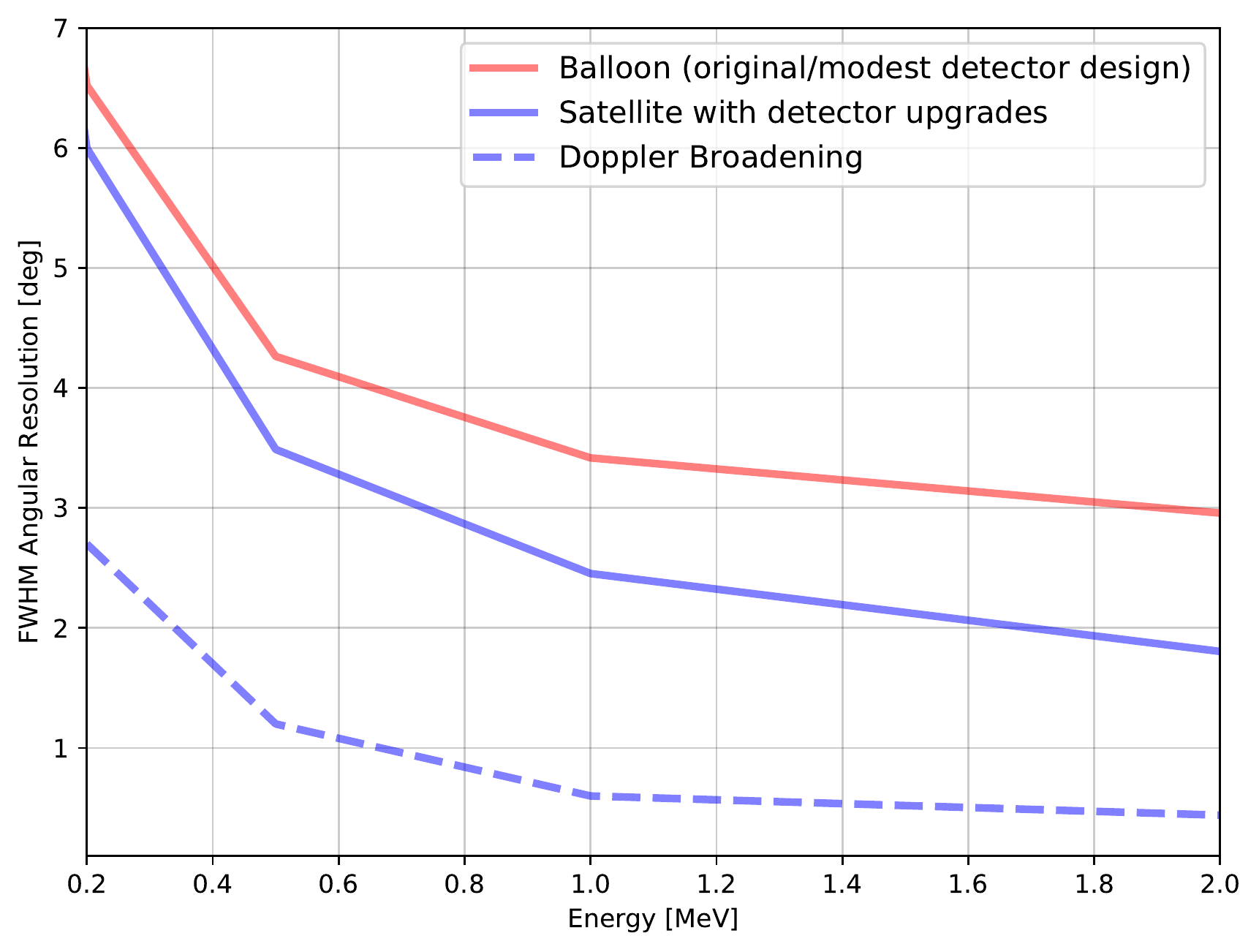}
\end{center}
\caption{The FWHM angular resolution in GRAMS compared with the uncertainty due to the Doppler broadening effect in the Compton scattering.}
\label{fig:Angular}
\end{figure}

The angular resolution $\sigma_{\theta}$ for Compton scattering events can be estimated as below \citep{Odaka2007}.

\begin{align*}
\sigma_{\theta}^2 &=  \delta\theta_E^2 + \delta \theta_r^2 + \delta\theta_{DB}^2 \\
&\delta\theta_E \simeq \frac{\sigma_E}{E}\\
&\delta\theta_r = \frac{\sigma_r}{d}\\
\end{align*}
where $\sigma_E$ is the energy resolution at $E$, $\sigma_r$ is the spatial resolution of the LArTPC, $d$ is the distance between the Compton scattering(s) and the photo-absorption, and $\delta\theta_{DB}$ is the uncertainty due to the Doppler broadening effect in the Compton scattering \citep{Boggs2001,Zoglauer2003}. The energy resolution of LArTPC is estimated from measurements by other experiments deploying similar detectors; DarkSide-50 and nEXO measured $\sigma_E \sim 5\%$ at 41.5 keV and $\sim 1\%$ at 2.5 MeV, respectively \citep{Agnes2015,Albert2018}. 

\begin{align*}
\sigma_E^2 &= \Delta E_{s}^2+\Delta E_{e}^2\\
&\frac{\Delta E_{s}}{E} \simeq \frac{1\%}{\sqrt{E \text{ (MeV)}/2.5}}
\end{align*}
where $\Delta E_{e}$ is the noise contribution from the electronics ($\sim 5$ keV) and $\Delta E_{s}$ is the statistical uncertainty for scintillation light and ionization electrons.

Figure \ref{fig:Angular} shows the Full Width at Half Maximum (FWHM, $2.35 \sigma_{\theta}$) of the angular resolution in GRAMS compared with the uncertainty due to the Doppler broadening effect in the Compton scattering. The angular resolution could be significantly improved with detector upgrades in the satellite mission, using a fine pitch of anode wires and an additional calorimeter as mentioned above. 

\subsection{Sensitivity}

\begin{figure}[t!]
\begin{center} 
\includegraphics*[width=8.0cm]{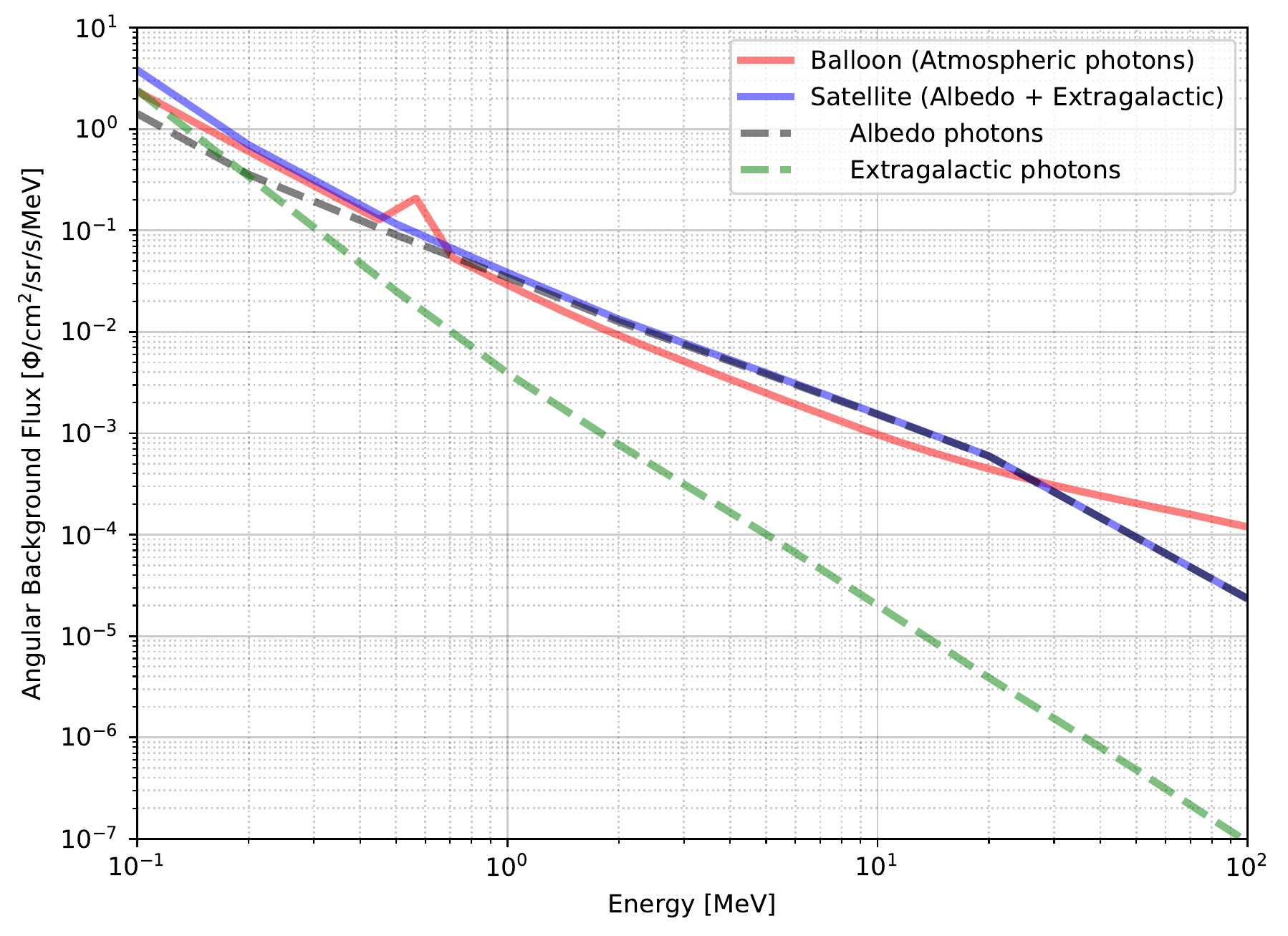}
\end{center}
\caption{The expected background fluxes for the balloon experiment (atmospheric photons) and the satellite mission in a low-earth orbit (Albedo and Extragalactic photons) \citep{Cumani2019}.}
\label{fig:Bkg}
\end{figure}

The gamma-ray continuum sensitivity ($S_{cnt,k}$ [$ph/cm^2/MeV/s$]) in GRAMS was estimated assuming a background count limited observation. 

\begin{align*}
S_{cnt,k}(E) \simeq k\sqrt{\frac{\Phi_{B} \Delta\Omega}{A_{eff} T_{eff} \Delta E}}
\end{align*}
where $k$ is the significance level of the source detection, $\Phi_{B}$ [$ph/cm^2/MeV/sr/s$] is the background flux, $T_{eff}$ is the effective observation time, $\Delta E$ [$MeV$] is the energy bandwidth around $E$ ($\Delta E = 0.5E$), and $\Delta \Omega$ [$sr$] is the solid angle corresponding to the angular resolution. Figure \ref{fig:Bkg} shows the expected background fluxes for the balloon experiment (atmospheric photons obtained from EXPACS\footnote{EXcel-based Program for calculating Atmospheric Cosmic ray Spectrum (EXPACS) instantaneously calculates terrestrial cosmic ray fluxes of neutrons, protons, and ions with charge up to 28 (Ni) as well as muons, electrons, positrons, and photons nearly anytime and anywhere in the Earth's atmosphere (https://phits.jaea.go.jp/expacs/).}) and the satellite mission in a low-earth orbit (Albedo and Extragalactic photons) \citep{Cumani2019}. Neutrons can also be the source of background in MeV gamma-ray observations, but LArTPC detectors are capable of distinguishing gamma rays from neutrons based on the pulse shape of the scintillation signal as has been demonstrated in dark matter search experiments \citep{Agnes2015,Boulay2006}. The LArTPC detector is made of relatively low-z material and, unlike INTEGRAL \citep{Jean2000,Roques2003,Weidenspointner2003,Diehl2017}, we may not see more than ten gamma-ray lines from the activation in LArTPC, based on the measurement with a neutron beam as well as production rates of cosmogenic-induced isotopes with the ACTIVIA program \citep{Gray1965,Back2008}. The detailed background simulations will be done in future studies, especially taking into account the instrumental background for the satellite mission due to high-energy cosmic rays. We will also evaluate the additional background rejection capability with detector upgrades in the satellite mission.

\begin{figure*}[t!]
\begin{center} 
\includegraphics*[width=12cm]{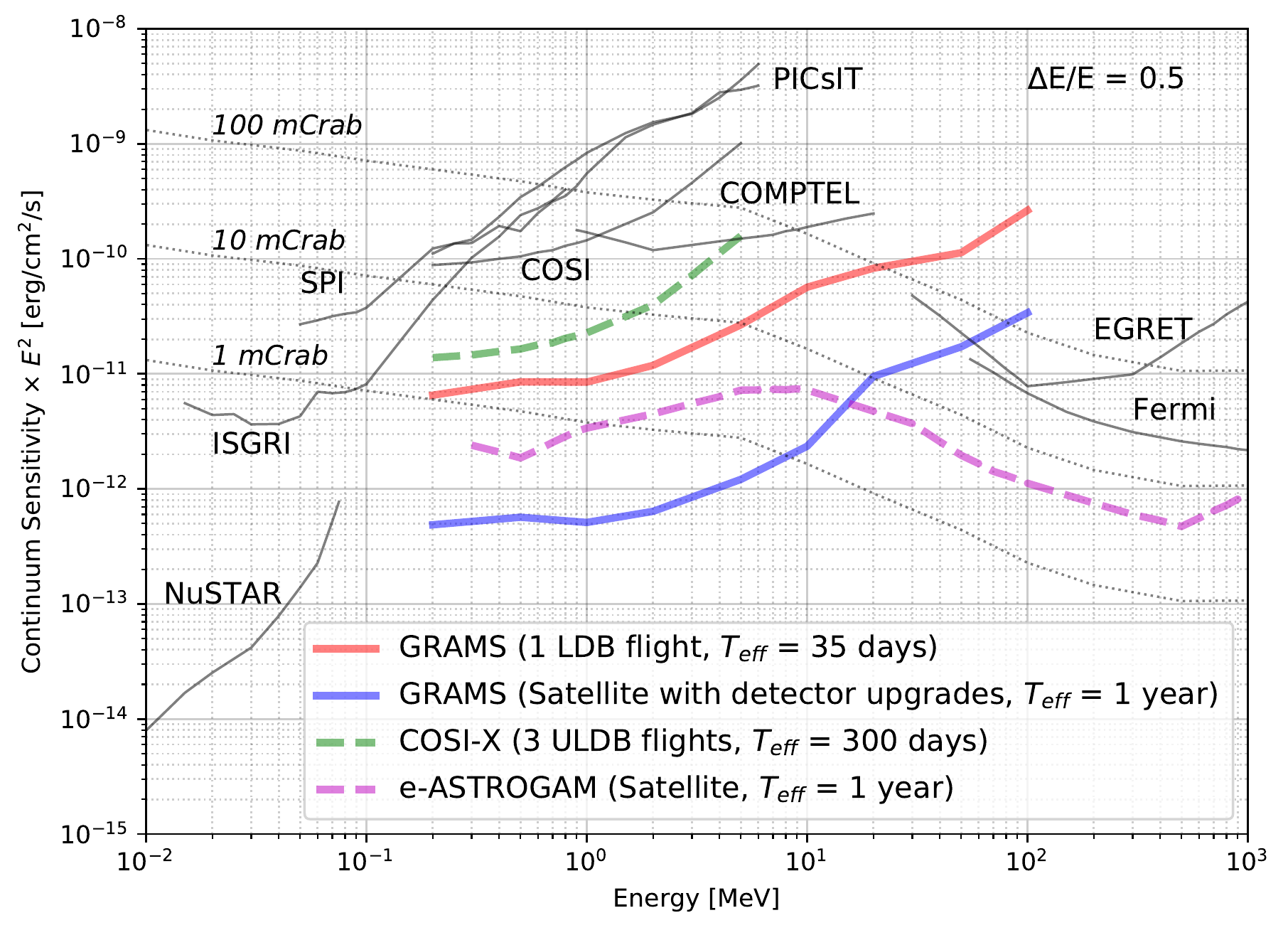}
\end{center}
\caption{The continuum gamma-ray sensitivities at a 3$\sigma$ confidence level for the GRAMS balloon experiment (one LDB flight, $T_{eff}$ = 35 days) and a possible satellite mission with detector upgrades ($T_{eff} = 1$ year) compared to the sensitivities for previous and future experiments, including SPI (SPectrometer for INTEGRAL), PICsIT (PIxelated CsI Telescope) and ISGRI (INTEGRAL Soft Gamma-Ray Imager) on INTEGRAL.  Black dashed lines represent the flux levels of 1-100 mCrab \citep{Takahashi2012,De2017}.}
\label{fig:S_GR}
\end{figure*}

Figure \ref{fig:S_GR} shows the GRAMS gamma-ray continuum sensitivities at a 3$\sigma$ confidence level for one LDB flight ($T_{eff}$ = 35 days) and the satellite mission with detector upgrades ($T_{eff} = 1$ year). Flux levels for 1-100 mCrab are shown for reference. GRAMS would be able to extensively explore gamma rays in the MeV energy domain. In particular, the sensitivity for a single LDB flight could be an order of magnitude improved compared to previous experiments\footnote{COSI collaboration website (The Compton Spectrometer and Imager, http://cosi.ssl.berkeley.edu)} \citep{Takahashi2012} and a few times better than the sensitivities for the future COSI-X mission\footnote{The sensitivity was estimated based on the observation time and improved angular resolution, compared to the COSI sensitivity.} with three Ultra-Long-Duration Balloon (ULDB) flights (3 $\times$ 100 days). The sensitivity for the GRAMS satellite mission could be a few times better than the sensitivity for e-ASTROGAM at $E <$ 10 MeV. e-ASTROGAM requires $\sim$50 layers of double-sided silicon strip detectors with a large number ($\sim$10$^6$) of channels/electronics \citep{De2017} while GRAMS uses a cost-effective LArTPC detector with a smaller number ($\sim$10$^4$ with a fine pitch of anode wires) of channels/electronics.

\begin{table*}[h!]
\begin{center}
\begin{tabular}{|c|c|c|c|}
\hline
Sensitivity [ph/cm$^2$/s] & GRAMS Balloon (Satellite) & SPI/INTEGRAL& Improvement Factor \tabularnewline \hline
$e^+$ (511 keV) &  2.9 $\times$ 10$^{-6}$ (6.3 $\times$ 10$^{-7}$) & 5.0 $\times 10^{-5}$ & $\sim$15 ($\sim$80) \tabularnewline \hline
$^{126}$Sn (666/695 keV) & 2.1 $\times$ 10$^{-6}$ (4.2 $\times$ 10$^{-7}$) & $\sim$2 $\times 10^{-5}$ & $\sim$10 ($\sim$50) \tabularnewline \hline
$^{56}$Co (847 keV) & 1.4 $\times$ 10$^{-6}$ (2.7 $\times$ 10$^{-7}$) & $\sim$2 $\times 10^{-5}$ & $\sim$15 ($\sim$75) \tabularnewline \hline
$^{44}$Ti (1157 keV) & 1.0 $\times$ 10$^{-6}$ (1.9 $\times$ 10$^{-7}$) & $\sim$2 $\times 10^{-5}$ & $\sim$20 ($\sim$110) \tabularnewline \hline
$^{60}$Fe (1173 keV) & 1.0 $\times$ 10$^{-6}$ (1.9 $\times$ 10$^{-7}$) & $\sim$2 $\times 10^{-5}$ & $\sim$20 ($\sim$110) \tabularnewline \hline
$^{60}$Fe (1333 keV) & 9.1 $\times$ 10$^{-7}$ (1.7 $\times$ 10$^{-7}$) & $\sim$2 $\times 10^{-5}$ & $\sim$20 ($\sim$120) \tabularnewline \hline
$^{26}$Al (1809 keV) & 7.2 $\times$ 10$^{-7}$ (1.3 $\times$ 10$^{-7}$) & 2.5 $\times 10^{-5}$ & $\sim$35 ($\sim$190) \tabularnewline \hline
$^{2}$H (2223 keV) & 6.4 $\times$ 10$^{-7}$ (1.1 $\times$ 10$^{-7}$) & $\sim$2 $\times 10^{-5}$ & $\sim$30 ($\sim$180) \tabularnewline \hline
$^{12}$C* (4438 keV) & 4.9 $\times$ 10$^{-7}$ (7.3 $\times$ 10$^{-8}$) & $\sim$1 $\times 10^{-5}$ & $\sim$20 ($\sim$140) \tabularnewline \hline
\end{tabular}
\caption{The GRAMS line sensitivity for the balloon (satellite) experiment (3$\sigma$, $T_{eff} = 10^6$ s) to positron annihilation and radioactive isotopes compared with SPI/INTEGRAL (3$\sigma$, $T_{eff} = 10^6$ s).}
\label{tbl:S_Line}
\end{center}
\end{table*}

The sensitivity to gamma-ray lines from positron annihilation and radioactive isotopes ($S_{line,k}$ [$ph/cm^2/s$]) was also estimated. The bandwidth ($\Delta E$) was chosen as 2$\sigma_E$ at each energy. 

\begin{align*}
S_{line,k}(E) &\simeq k\sqrt{\frac{\Phi_{B} \Delta\Omega \Delta E}{A_{eff} T_{eff}}}
\end{align*}

Table \ref{tbl:S_Line} shows the sensitivity for the balloon (satellite) experiment (3$\sigma$, $T_{eff} = 10^6$ s) to gamma-ray lines from positron annihilation (511 keV), $^{56}$Co (847 keV) from thermonuclear supernovae, $^{44}$Ti (1157 keV) from core-collapse supernovae, $^{60}$Fe (1173 keV and 1333 keV) from core-collapse supernovae, $^{26}$Al (1809 keV) from collapse supernovae or massive stars, $^2$H (2223 keV) from neutron capture by protons, and $^{12}$C* (4438 keV) from cosmic ray interactions \citep{Knodlseder2001,Diehl2013}. GRAMS could also detect gamma-ray lines from Galactic neutron star merger remnants (666 keV and 695 keV from $^{126}$Sn) \citep{Wu2019}. With this larger effective area, GRAMS would have more than an order of magnitude improved sensitivity compared with SPI/INTEGRAL \citep{Jean2000,Roques2003,Weidenspointner2003,Diehl2017}. 



\section{Antimatter Survey} \label{sec:AM}

\begin{figure*}[t!]
\begin{center} 
\includegraphics*[width=16cm]{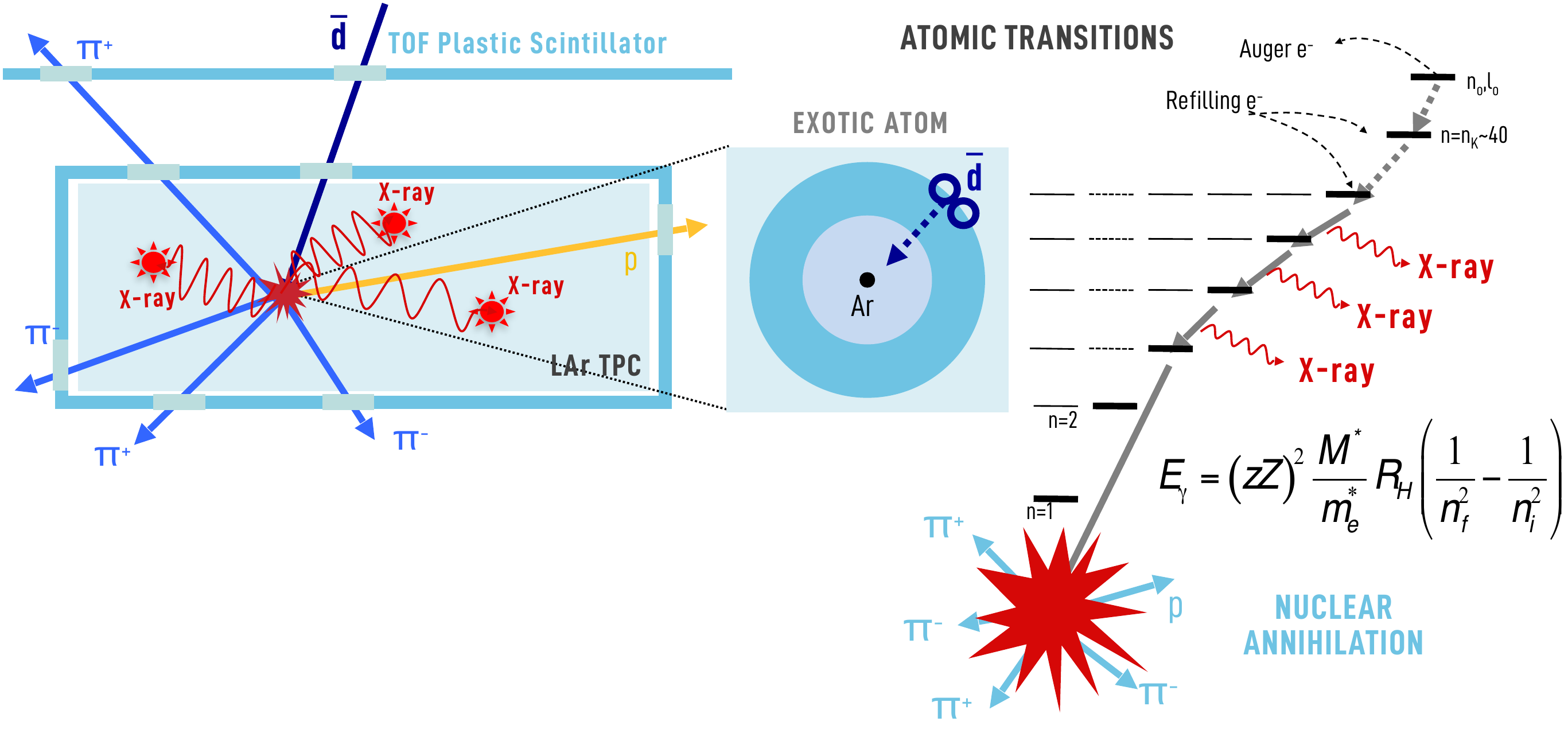}
\end{center}
\caption{The GRAMS antimatter detection technique. The stopped antimatter forms an excited exotic atom that decays and emits atomic X-rays and annihilation products (pions and protons). The atomic X-rays, pion/proton multiplicities, and the stopping depth in the LArTPC provide the particle identification capability.}
\label{fig:Detection_AM}
\end{figure*}

\subsection{Detection Concept}

The GRAMS antimatter survey involves capturing an antiparticle in a target material with subsequent formation and decay of an exotic atom, similar to the GAPS project \citep{Mori2002,Aramaki2013,Aramaki2016}. The TOF plastic scintillators measure the velocity and angle of the incoming antiparticle. The antiparticle slows down and stops inside the LArTPC detector, where it forms an excited exotic atom with an argon atom. Then, the exotic atom emits atomic X-rays as it de-excites. The energy of the atomic X-ray depends on the mass and atomic number of the antiparticle and the target atom. The two highest atomic X-rays for an exotic atom with an antiproton and an argon atom are 58 keV and 97 keV, and 74 keV and 114 keV with an antideuteron and an argon atom. At the end of the atomic cascade, the antiparticle subsequently annihilates in the nucleus with the emission of pions and protons. The number of pions and protons produced (pion/proton multiplicity) is roughly proportional to the number of antinucleons in the incoming antiparticle. This pion/proton multiplicity can also be used to identify the incoming antiparticle. 

\begin{figure}[b!]
\begin{center} 
\includegraphics*[width=7.5cm]{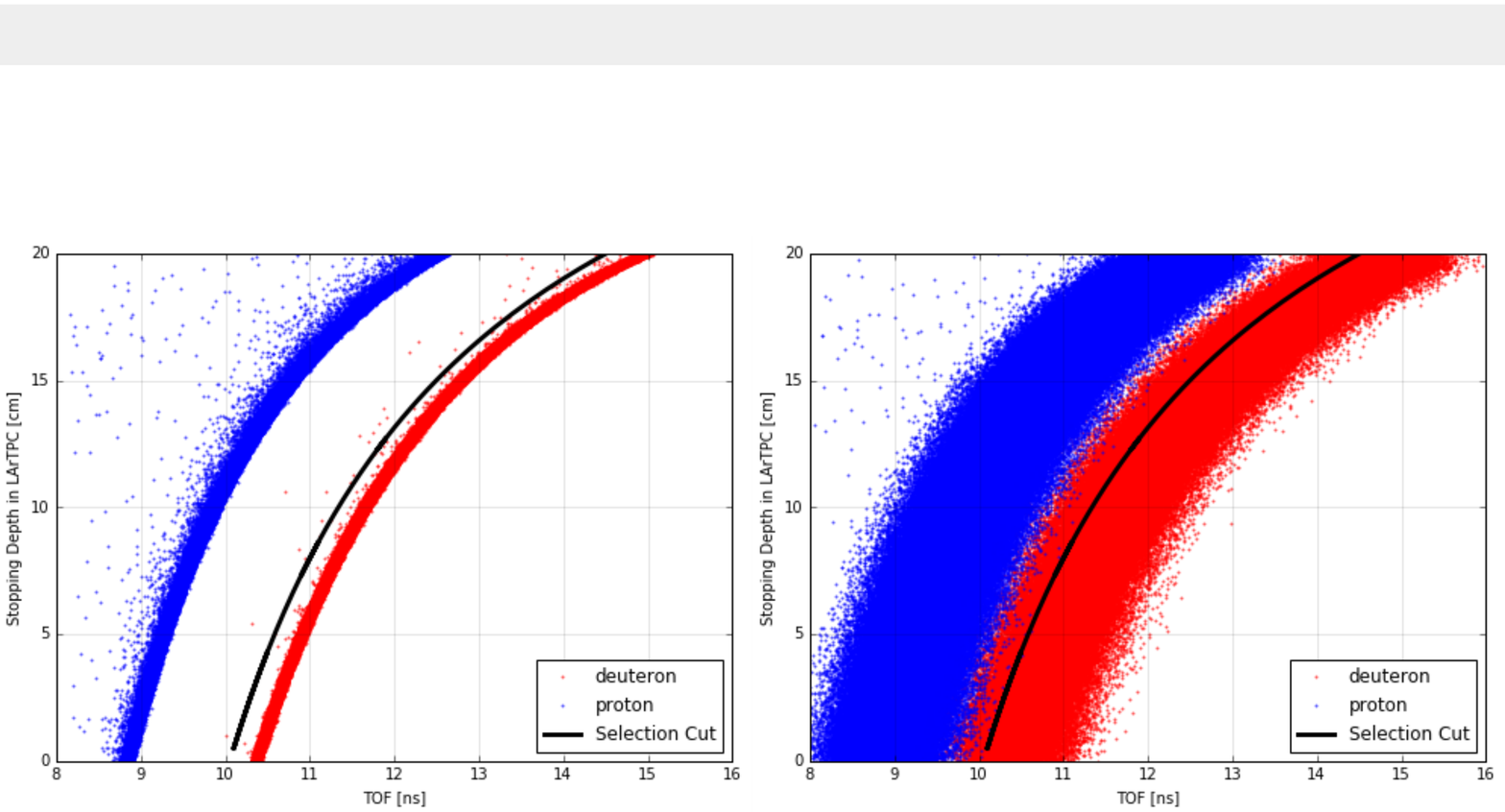}
\end{center}
\caption{The simulation result for the stopping range of protons and deuterons taking into account the position resolution of $\sim$3 mm in the LArTPC detector and the timing resolution of $\sim$0.4 ns in the TOF system. The solid black line represents the selection cut.}
\label{fig:Depth}
\end{figure}

\begin{figure*}[t!]
\begin{center} 
\includegraphics*[width=16cm]{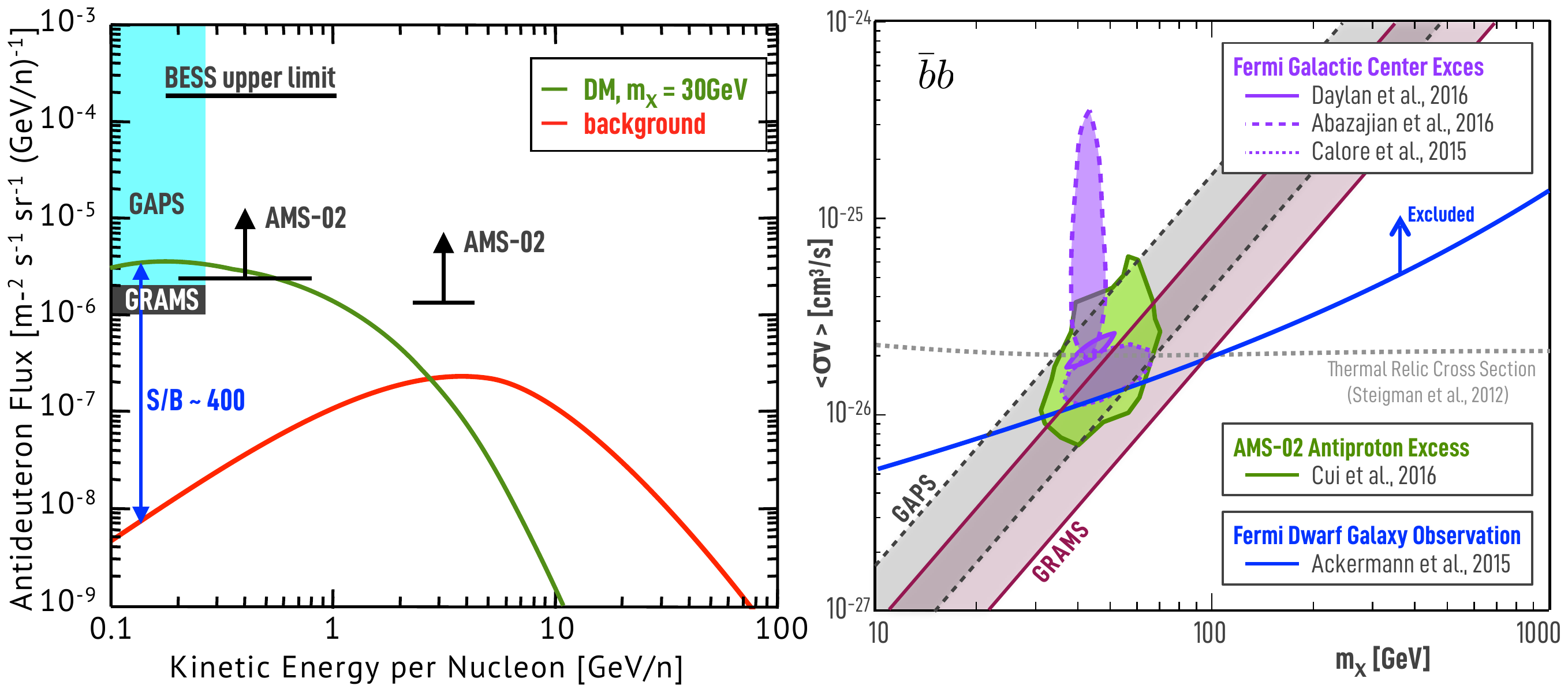}
\end{center}
\caption{The left figure shows the antideuteron sensitivities for GRAMS (three LDB flights from Antarctica, 105 days) and other experiments together with the predicted antideuteron fluxes from dark matter annihilation (primary) and cosmic ray interactions (secondary) \citep{Donato2000,Fuke2005,Aramaki2016,Ong2017,Donato2008,Ibarra2013}. The right figure shows the parameter space for the possible dark matter signatures suggested by Fermi and AMS-02 \citep{Calore2015,Daylan2016,Abazajian2016,Ackermann2017,Cui2016}, where GRAMS and GAPS sensitivities are overlaid with an uncertainty of the antideuteron production model \citep{Korsmeier2018,Fornengo2013,Acharya2018}. GAPS is an antideuteron search experiment using the exotic atom technique that is currently under construction for a science flight with a first opporunity in the 2020 austral summer.}
\label{fig:S_AM}
\end{figure*} 

Additionally, GRAMS can measure the stopping range of the incoming antiparticles with an excellent position resolution in the LArTPC. Figure \ref{fig:Depth} shows the simulation results for the stopping range study, assuming the position resolution of $\sim$3 mm in the LArTPC detector and the timing resolution of $\sim$0.4 ns in the TOF system. Here, protons and deuterons were used instead of antiprotons and antideuterons to simplify the physics process. The solid black line represents the selection cut while the acceptances (probabilities for events in the right side of the selection cut) are 2.7e-6 for antiprotons and 0.74 for antideuterons. These particle identification techniques (atomic X-rays, pion/proton multiplicities, and stopping range) will allow us to distinguish incoming antideuterons from others. For example, the TOF system will select slow incoming particles, which allows us to identify and reject the cosmic ray protons since low-energy protons may not be able to produce relativistic pions as seen in the nuclear annihilation. GRAMS could also identify cosmic ray species based on the combination of the TOF, dE/dX energy deposit, and the stopping range, which can be used for the solar modulation study. The detection concept and particle identification techniques were validated and demonstrated in the accelerator test with an antiproton beam as well as the prototype flight for the GAPS experiment \citep{Aramaki2013,Doetinchem2014,Mognet2014,Fuke2014}. The expected mimic/background events could be as small as 0.01 during the LDB flight. 

\subsection{Sensitivity}  \label{sec:AM_Sensitivity}

The left panel of Figure \ref{fig:S_AM} shows the GRAMS antideuteron sensitivity (three LDB flights from Antarctica, 105 days) together with the predicted antideuteron fluxes from dark matter annihilation (primary) and cosmic ray interactions (secondary/background). The primary antideuteron flux can be more than two orders of magnitude larger than the secondary flux at low-energy \citep{Donato2000,Donato2008,Ibarra2013}. Therefore, the GRAMS antideuteron measurement is essentially a background-free dark matter search and provides stringent constraints on a variety of dark matter models. The GRAMS antideuteron sensitivity could be a few times better than the current generation experiments, GAPS (three LDB flights, 105 days) and AMS-02\footnote{The AMS-02 sensitivity is a strict upper limit since it is based on the superconducting magnet, rather than the permanent magnet used in the actual flight.} (five years of observation)\citep{Aramaki2016,Ong2017}, and more than two orders of magnitude better than the current upper limit obtained by BESS \citep{Fuke2005}. 


The recent results of the Fermi gamma-ray observation from the Galactic Center region and spheroidal dwarf galaxies as well as the AMS-02 antiparticle measurements indicate the detection of possible dark matter signatures \citep{Calore2015,Daylan2016,Abazajian2016,Cui2016,Ackermann2017}. The right panel of Figure \ref{fig:S_AM} shows the parameter space for the possible dark matter signatures suggested by Fermi and AMS-02, where GRAMS and GAPS sensitivities are overlaid with an uncertainty of the antideuteron production model \citep{Korsmeier2018,Fornengo2013,Acharya2018}. With given uncertainties, GAPS would be able to deeply cover the parameter space for these dark matter signatures. GRAMS could fully investigate the parameter space and could obtain some crude spectrum information for antideuterons produced by dark matter annihilation or decay. The GRAMS detector is cost-effective and relatively simple to configure considering that the detector is a single layer design and can be cooled down with a commercially available cryocooler. Because the LArTPC system can be easily and cost-effectively replaced if damaged on experiment recovery, this would allow for multiple balloon flights per a few years and further improve the sensitivities for both MeV gamma-ray and antimatter measurements.


\section{Conclusion}

GRAMS will be the first balloon experiment optimized for both MeV gamma-ray observations and an indirect dark matter search with a LArTPC detector. With a cost-effective, large-scale LArTPC detector, GRAMS can have more than an order of magnitude improved sensitivity to MeV gamma rays and a few times better sensitivity to antideuterons compared to the previous and current ongoing experiments. GRAMS is moving towards the next generation experiment and currently is in the development phase. A reasonable development timescale for GRAMS would put its first balloon flight just beyond the completion of the three flight GAPS mission and the future COSI-X mission.

\section{Acknowledgments}

This work is supported and funded by Department of Energy (DE-AC02-76SF00515) and NASA APRA Grants (NNX09AC16G). We would like to thank Tom Shutt, Kazuhiro Terao, Greg Madejski, and Chris Stanford at SLAC/Stanford University as well as Yoshiyuki Inoue at RIKEN for helpful discussions. We would also like to thank the GAPS collaboration for useful suggestions and discussions.

\bibliographystyle{model1-num-names}

\bibliography{refs}






\end{document}